# Performance of UAV-based Cell-free mMIMO ISAC Networks: Tethered vs. Mobile

Xavier A. Flores Cabezas*[†][§]; Isabella W. G. da Silva*[‡][§]; and Markku Juntti*
*Centre for Wireless Communications (CWC), University of Oulu, Finland,
[†]Interdisciplinary Centre for Security, Reliability and Trust (SnT), University of Luxembourg, Luxembourg,
[‡]Centre for Wireless Innovation (CWI), Queen's University Belfast, U.K.
e-mails: alejandro.flores@uni.lu, iwgdasilva01@qub.ac.uk, markku.juntti@oulu.fi
[§]Authors contributed equally

*Abstract*—The employment of unmanned aerial vehicles (UAVs) aligned with multistatic sensing in integrated sensing and communication (ISAC) systems can provide remarkable performance gains in sensing, by taking advantage of the cell-free massive multiple-input multiple-output (mMIMO) architecture. Under these considerations, in this paper, the achievable sensing signal-to-noise-plus-interference ratio (SINR) of a cell-free mMIMO ISAC UAV-based network is evaluated for two different deployments of UAVs, namely, mobile and tethered. In both scenarios, a transmit precoder that jointly optimizes the sensing and communication requirements subjected to power constraints is designed. Specifically, for the scenario with mobile UAVs, beyond the transmit precoding, we also optimize the position of the transmit UAVs through particle swarm optimization (PSO). The results show that, although tethered UAVs have a more efficient power allocation, the proposed position control algorithm for the mobile UAVs can achieve a superior gain in terms of sensing SINR.

## I. INTRODUCTION

The sixth generation (6G) of wireless communications is expected to enable reliable and virtually limitless connectivity under different requirements of data rate, latency, and energy efficiency. Unmanned aerial vehicles (UAVs) enhanced connectivity can play an important role for 6G networks given their deployment flexibility, enabling several applications as support of wireless communications networks, environmental monitoring, and real-time surveillance [1].

There are several ways to classify UAVs, one of them being tethered or mobile UAVs. In the former, a UAV is connected to a ground station or terrestrial platform via a tethered cable, which transmits data and power from the ground to the air devices. The main idea of tethered UAVs is to overcome the limitations in terms of battery power, thus improving the connectivity of the devices. Moreover, given its inherently superior aerial safety, tethered UAVs may be allowed to operate in more populated areas [2], [3]. Specifically, in [3], different tethered UAV deployments are evaluated in terms of coverage capability and energy efficiency. The results demonstrated that, in comparison to fixed BSs, tethered UAVs can significantly improve the coverage capability. Mobile UAVs, on the other hand, are not connected via cables and, thus, depend on the battery power contained in the device. However, the range of mobility of the devices can be exploited to enhance the performance as in [4] and [5]. In particular, in [4], Qiu *et al.* propose a joint UAV placement, resource allocation, and user association design to improve the user throughput under backhaul constraints. Besides, in [5], the UAVs trajectory and the coordinated multi-point (CoMP) transmission are optimized in terms of the sum rate of ground users (UEs), with the results indicating a significant improvement in sum rate in comparison to static trajectory schemes.

In addition to connectivity requirement, 6G systems are expected to provide sensing capabilities. The idea is that, it will be possible to explore the reflections of the signals to extract information related to the physical space and to predict the movement of nodes. On this, integrated sensing and communications (ISAC) emerges as a promising technology, allowing the efficient usage of spectrum resources and support of several 6G use cases, such as vehicle-to-everything and smart homes [6]. Particularly, ISAC-enabled UAV networks can provide an additional degree of freedom for design and optimization [7], [8]. For instance, in [7], the radar echoes from an eavesdropper and a legitimate UE received by the UAV are used to optimize the UAV trajectory in terms of the real-time secrecy rate. In [8], the UAV maneuvering and transmit beamforming are designed to maximize the weighted sum-rate of communication users under beampattern gain requirements. It is valid to point out that the previous works, and most of the literature in ISAC considers monostatic setups, that is, with co-located transmitter and receiver for sensing, which requires full-duplex capability from the sensing transmitter/receiver. Accordingly, multistatic deployments with non-colocated transmitters and receivers are capable of offering a diversity gain, and differently from the monostatic design, do not require full-duplex capability from the nodes [9], [10]. Both [9] and [10] demonstrate that multistatic deployments, based on cell-free massive multiple-input multiple-output (mMIMO) ISAC designs require less transmit power to attain an accurate detection of the target.

Accordingly, recognizing the benefits of UAVs and multistatic sensing in ISAC systems, in this paper, we investigate a UAV-based cell-free mMIMO ISAC network with multiple UEs and a single point-like target in terms of the sensing signal-to-noise-plus-interference ratio (SINR). In this scenario, we compare the performance of tethered and mobile UAVs. The main contributions of this work are as follows: i) We propose a transmit precoding design to maximize the sensing

SINR constrained to a minimum SINR for the UEs and maximum transmit power. For the scenario with mobile UAVs, backhaul requirements are also considered. ii) For the scenario with mobile UAVs, we propose a block coordinate descent (BCD) method to optimize the position of the transmit UAVs and the transmit precoding. The former is based on a particle swarm optimization (PSO) algorithm and employs sub-optimal designs of the transmit precoding to define the best positions of the UAVs.

*Notation.* Throughout this paper, bold upper-case letters denote matrices whereas bold lower-case letters denote vectors; $(\cdot)^T$ and $(\cdot)^H$ stand for the matrix transpose and Hermitian transpose, respectively; $\mathbf{I}$ is the identity matrix; $||\cdot||$ and $|\cdot|$ are the Euclidean-norm and the absolute value operator; $d_{a,b}$ stands for the Euclidian distance between $a$ and $b$.

## II. SYSTEM MODEL

### A. Mobile UAVs

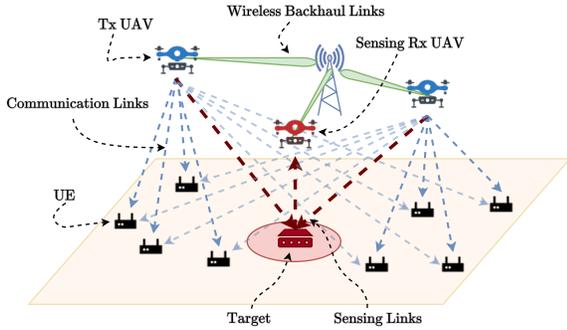

Fig. 1. UAV-assisted ISAC network with wireless backhaul system model.

As illustrated in Fig. 1, the considered downlink system comprises $N_{\text{Tx}}$ transmit UAVs working in a cell-free mMIMO manner, that is, jointly serving $N_{\text{ue}}$ single-antenna ground UEs and sensing a point-like target. Simultaneously, one Rx UAV acts as a sensing receiver. All the UAVs are equipped with two antenna arrays, namely a square uniform planar array (UPA) (mounted facing downward), to transmit communications data and sensing signals, and a uniform linear array (ULA) (mounted horizontally) to communicate with a ground BS via a wireless backhaul link. The UPAs and the ULAs are equipped with $M_{\text{U}}$ and $M_{\text{UB}}$ half-wavelength-spaced isotropic antenna elements, respectively. Moreover, the BS has a ULA equipped with $M_{\text{BS}}$ half-wavelength-spaced isotropic antenna elements. We further assume that the processing is made in a centralized manner, nodes are synchronized, and the fronthaul and backhaul links operate in different frequency bands as in [4]. Thus, at time instance $n$, the received signal at the $k$th transmit UAV from the BS is

$$y_k[n] = \mathbf{u}_k^H \left( \sum_{k=1}^{N_{\text{Tx}}} \mathbf{H}_{b,k} \mathbf{w}_{b,k} s_{b,k}[n] + \mathbf{n}_{b,k}[n] \right), \quad (1)$$

where $\mathbf{u}_k \in \mathbb{C}^{M_{\text{UB}}}$ is the receive beamformer, $s_{b,k}[n]$ and $\mathbf{w}_{b,k} \in \mathbb{C}^{M_{\text{BS}}}$ are the transmitted symbol and the transmit precoder vector by the BS to UAV $k$ at time $n$, respectively, and $\mathbf{n}_{b,k}$ is the zero-mean additive white Gaussian noise (AWGN) component at the $k$th transmit UAV with covariance $\sigma_k^2 \mathbf{I}$. Furthermore, the backhaul links are assumed to operate in the millimeter wave (mmWave) band so that the channel matrix between the BS and transmit UAV $k$, $\mathbf{H}_{b,k} \in \mathbb{C}^{M_{\text{UB}} \times M_{\text{BS}}}$ undergoes the Saleh-Valenzuela channel model, written as [11]

$$\mathbf{H}_{b,k} = \sqrt{\frac{M_{\text{UB}} M_{\text{BS}}}{L_{b,k}}} \sum_{p=0}^{L_{b,k}} g_{b,k}^p \mathbf{a}(\phi_{b,k}) \mathbf{a}^H(\varphi_{b,k}), \quad (2)$$

where $L_{b,k}$ is the number of resolvable paths, $\phi_{b,k}$ is the angle of arrival (AoA) at UAV $k$, and $\varphi_{b,k}$ is the angle of departure (AoD) at the BS. Accordingly, $\mathbf{a}(\phi_{b,k}) \in \mathbb{C}^{M_{\text{UB}}}$ and $\mathbf{a}(\varphi_{b,k}) \in \mathbb{C}^{M_{\text{BS}}}$ are antenna array steering vectors. Moreover, $g_{b,k}^p \sim \mathcal{CN}(0, 10^{-\alpha_{b,k}^p/10})$, $p=0,\ldots,L_{b,k}$, is the complex channel gain of the path. Precisely, $g_{b,k}^0$ stands for the line-of-sight (LoS) path, and $g_{b,k}^p$, $p \in \{1, L_{b,k}\}$ is the $p$th non-LoS (NLoS) path. Finally, $\alpha_{b,k}^p$ is the path loss, modeled as

$$\alpha_{b,k}^p = a_i + 10b_i \log(d_{b,k}) + \mu_i, i \in \{\text{LoS}, \text{NLoS}\}, \quad (3)$$

where $a_i$, $b_i$ and $\mu_i$ are constants. Accordingly, the SINR at UAV $k$ is given by

$$\gamma_k = \frac{|\mathbf{u}_k^H \mathbf{H}_{b,k} \mathbf{w}_{b,k}|^2}{\sum_{\substack{l=1 \\ l \neq k}}^{N_{\text{Tx}}} |\mathbf{u}_k^H \mathbf{H}_{b,k} \mathbf{w}_{b,l}|^2 + \sigma_k^2 ||\mathbf{u}_k||^2}. \quad (4)$$

Next, the signal transmitted by the $k$th UAV at time $n$, is a weighted sum of communications and sensing signals as

$$\mathbf{x}_k[n] = \sum_{j=1}^{N_{\text{ue}}} \mathbf{w}_{j,k} s_j[n] + \mathbf{w}_{t,k} s_t[n] = \mathbf{W}_k \mathbf{s}[n] \in \mathbb{C}^{M_{\text{U}}}, \quad (5)$$

where $\mathbf{s}[n] = [s_1[n], \ldots, s_{N_{\text{ue}}}[n], s_t[n]]^T \in \mathbb{C}^{N_{\text{ue}}+1}$ contains the $N_{\text{ue}}$ parallel communications symbols intended to the UEs plus the sensing signal, which is independent of UEs' data signals. Also, $\mathbf{w}_{j,k}, \mathbf{w}_{t,k} \in \mathbb{C}^{M_{\text{U}}}$ are the transmit precoder vectors of the $k$th UAV for the $j$th UE and for the sensing of the target, respectively. Thus, the received signal at the $j$th UE, at time $n$, is given by

$$y_j[n] = \sum_{k=1}^{N_{\text{Tx}}} \mathbf{h}_{j,k} \mathbf{W}_k \mathbf{s}[n] + n_j[n], \quad (6)$$

where $\mathbf{h}_{j,k} \in \mathbb{C}^{M_{\text{U}}}$ is the air-to-ground (A2G) channel coefficient vector between UAV $k$ and UE $j$, given by $\mathbf{h}_{j,k} = \sqrt{\alpha_{j,k}^{-1}} \bar{\mathbf{h}}_{j,k}$. Where $\bar{\mathbf{h}}_{j,k}$ is the small-scale fading, modeled as a Rician fading channel as [12]

$$\bar{\mathbf{h}}_{j,k} = \sqrt{\frac{K_{j,k}}{K_{j,k}+1}} \bar{\mathbf{h}}_{j,k}^{\text{LoS}} + \sqrt{\frac{1}{K_{j,k}+1}} \bar{\mathbf{h}}_{j,k}^{\text{NLoS}}, \quad (7)$$

with $K_{j,k}$ being the Rician factor, computed as $A_1 e^{A_2 \theta_{j,k}}$, where $A_1$ and $A_2$ are constants, and $\theta_{j,k}$ is the corresponding elevation angle. Furthermore, $\alpha_{j,k}$ is the average path loss, given as [13]

$$\alpha_{j,k} = (2\pi \lambda_c d_{k,j})^\psi (P_{\text{LoS}} \eta_{\text{LoS}} + P_{\text{NLoS}} \eta_{\text{NLoS}}), \quad (8)$$

where $\psi$ is the path loss exponent, $\lambda_c$ is the carrier wavelength, and $\eta_i$, $i\in\{\text{LoS},\text{NLoS}\}$ is the attenuation factor for the LoS or NLoS link. Also, $P_{\text{LoS}}$ and $P_{\text{NLoS}}$ are the probabilities of LoS and NLoS connections given respectively by

$$P_{\text{LoS}} = \left(1+\varrho\exp\left(-\omega\left[\frac{180}{\pi}\tan^{-1}\left(\frac{z_k}{r_{k,j}}\right)-\varrho\right]\right)\right)^{-1}, \quad (9)$$

and $P_{\text{NLoS}} = 1 - P_{\text{LoS}}$, where $\varrho$ and $\omega$ are constants, $z_k$ is the altitude of UAV $k$, and $r_{k,j}$ is the distance from UE $j$ to the ground projection of UAV $k$. In addition, $n_j$ is the noise component, modeled as AWGN with variance $\sigma_j^2$. For simplicity, we consider $\mathbf{h}_j = [\mathbf{h}_{j,1},\ldots,\mathbf{h}_{j,N_{\text{Tx}}}]^T$, and $\tilde{\mathbf{W}} = [\mathbf{w}_1,\ldots,\mathbf{w}_{N_{\text{ue}}},\mathbf{w}_t]$, with $\mathbf{w}_j = [\mathbf{w}_{j,1}^T,\ldots,\mathbf{w}_{j,N_{\text{Tx}}}^T]^T$ $\forall j \in \{1,N_{\text{ue}}\}$ and $\mathbf{w}_t = [\mathbf{w}_{t,1}^T,\ldots,\mathbf{w}_{t,N_{\text{Tx}}}^T]^T$. Thus, the SINR at the $j$th UE is written as

$$\gamma_j = \frac{|\mathbf{h}_j\mathbf{w}_j|^2}{\sum_{\substack{l=1\\l\neq j}}^{N_{\text{ue}}}|\mathbf{h}_j\mathbf{w}_l|^2 + |\mathbf{h}_j\mathbf{w}_t|^2 + \sigma_j^2}. \quad (10)$$

On the other hand, the received signal at the receiver UAV, at time $n$, is given by

$$\mathbf{y}_r[n] = \sum_{k=1}^{N_{\text{Tx}}} \beta_{r,k}\sqrt{g_{r,k}}\mathbf{a}(\phi_{r,t},\theta_{r,t})\mathbf{a}^T(\phi_{k,t},\theta_{k,t})\mathbf{x}_k[n] + \mathbf{n}_r[n], \quad (11)$$

where $\beta_{r,k}$ is the bi-static unknown radar cross section (RCS) of the target through the reflection path from transmit UAV $k$ to the receiver UAV. Similar to [9], we assume that the RCS follows the Swerling-I model, with distribution $\beta_{r,k}\sim\mathcal{CN}(0,\sigma_{r,k}^2)$. Also, $\phi_{i,t}$ and $\theta_{i,t}$ are the azimuth and elevation angles from the target position to the $i$th UAV, with $i\in\{k=1,\ldots,N_{\text{Tx}},r\}$, respectively. Finally, $\mathbf{n}_r[n]\sim\mathcal{CN}(\mathbf{0},\sigma_r^2\mathbf{I}_{M_U})$ is the noise component at the receiver UAV, and $g_{r,k}$ is the channel gain of the path between the $k$th transmit UAV to the target and from the target to the receiver UAV, computed as [9]

$$g_{r,k} = \frac{\lambda_c^2}{(4\pi)^3 d_{t,k}^2 d_{t,r}^2}, \forall k \in \{1,N_{\text{Tx}}\}. \quad (12)$$

Accordingly, neglecting the clutter caused by permanent or temporary objects, and assuming that for all $k$, $\beta_{r,k}$ are independent and share the same variance, $\sigma_{r,k}^2 = \sigma_{\text{rcs}}^2$, the sensing SINR is written as

$$\gamma_t = \zeta\sum_{n=1}^{N-1}\sum_{k=1}^{N_{\text{Tx}}}\frac{|\mathbf{s}^H[n]\mathbf{W}_k^H\mathbf{a}^*(\phi_{k,t},\theta_{k,t})|^2}{d_{k,t}^2}, \quad (13)$$

with $\zeta = \lambda_c^2\sigma_{\text{rcs}}/(N-1)(4\pi)^3\sigma_r^2 d_{r,t}^2$.

### B. Tethered UAVs

Fig. 2 illustrates a downlink system of an ISAC network assisted by tethered transmit and receive UAVs. Different from the system presented in Sec. II-A, all UAVs are fixed to the ground through tethers, and connected to the BS via a wired backhaul, which also provides power to the UAVs.

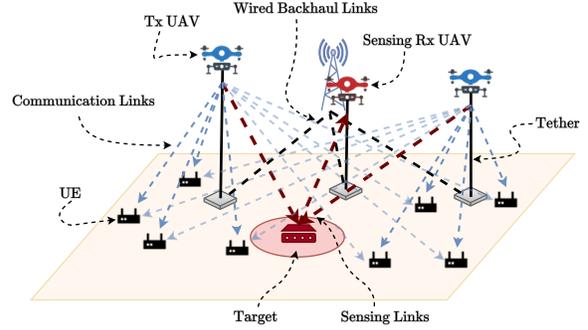

Fig. 2. Tethered UAV-assisted ISAC network system model

Besides, the processing is assumed centralized and fully synchronized. Thus, the transmit signal by the tethered UAVs, and the received by the UEs and sensing receiver, with their corresponding channels are analogous to those in Sec. II-A.

## III. MAXIMIZATION OF THE SENSING SINR

In this section, we present the strategies proposed to maximize the sensing SINR with mobile and tethered UAVs.

### A. Mobile UAVs

The goal is to define the best transmit and receive precoders $\mathbf{W} = [\tilde{\mathbf{W}},\mathbf{w}_b]$, $\mathbf{u}$, and transmit UAVs positions that maximize (12) under transmit power, communication and backhaul quality of service (QoS) constraints, which can be formulated as

$$\mathcal{P}_m : \max_{\mathbf{W},\mathbf{u},\mathbf{r}} \gamma_t \quad (14a)$$

$$\text{s.t. } \gamma_j \geq \Gamma, j \in \{1,N_{\text{ue}}\}, \quad (14b)$$

$$\min_k\{\log_2(\gamma_k+1)\} \geq N_{\text{ue}}\log_2(\Gamma+1), k \in \{1,N_{\text{Tx}}\}, \quad (14c)$$

$$||\mathbf{w}_b||^2 \leq P_b, \quad (14d)$$

$$\sum_{j=1}^{N_{\text{ue}}}||\mathbf{w}_{j,k}||^2 + ||\mathbf{w}_{t,k}||^2 \leq P_k, k \in \{1,N_{\text{Tx}}\} \quad (14e)$$

$$\mathbf{r}_{\min} \leq \mathbf{r}_k \leq \mathbf{r}_{\max}, k \in \{1,N_{\text{Tx}}\}, \quad (14f)$$

$$d_{k,i} \geq d_{\min}, \{k,i\} \in \{1,N_{\text{Tx}}\}, k \neq i, \quad (14g)$$

where $\Gamma$ is the communication SINR threshold, $P_b$ is the transmit power limit by the BS, and $P_k$ is the power limit by transmit UAV $k$, $\mathbf{r} = [\mathbf{r}_1,\ldots,\mathbf{r}_{N_{\text{Tx}}}]^T$, with $\mathbf{r}_k$ is set as the three-dimensional (3D) coordinate vector of UAV $k$, $\mathbf{r}_{\min} = [x_{\min},y_{\min},z_{\min}]$ and $\mathbf{r}_{\max} = [x_{\max},y_{\max},z_{\max}]$ are the minimal and maximal coordinates allowed for the UAVs to operate, and $d_{\min}$ is the minimal distance between any two UAVs to avoid collisions. To solve $\mathcal{P}_m$, a BCD method is employed. In BCD, the original optimization problem is divided into sub-problems that, at each iteration, are solved for a single block of variables while the others remain fixed. Accordingly, we consider two main blocks of optimization to maximize the sensing SINR with mobile UAVs, namely, the UAV position protocol and the precoder design detailed below.

*1) UAV Position Protocol:* Since the system is centrally coordinated, we consider that the transmit UAV position protocol is designed at the BS. Moreover, given that (14a) is neither convex nor concave regarding the position of the UAVs, and that the movement of the UAVs spans a continuous space, the optimization problem is computationally intractable. Thus, a solution to solve the position allocation problem is to resort to metaheuristic algorithms. On that, the PSO method is a good approach to determine the optimal position for the transmit UAV because it is robust for most optimization problems, easy to implement, has fewer parameters to adjust compared to similar optimization techniques, and converges to the global optimal in most cases. Precisely, the PSO method consists of several instances of the problem named particles being randomly initialized and iteratively updated until a convergence criterion is met [14]. In this case, we consider the utility function as the sensing SINR $\gamma_t$, and the particles as characterized by the positions of the UAVs $\mathbf{r}$ [1].

Furthermore, although the precoder design presented in Sec. III-A2 does not show a high complexity, it is computationally costly to compute for every particle at each step of the PSO algorithm. Accordingly, a zero-forcing (ZF) sub-optimal precoder for the UEs is considered. Assuming $\mathbf{H}_{k,\bar{j}} = [\mathbf{h}_{k,1}, ..., \mathbf{h}_{k,j-1}, \mathbf{h}_{k,j+1}, ..., \mathbf{h}_{k,N_{\text{ue}}}]$, the ZF precoder from UAV $k$ to UE $j$ is given by

$$\mathbf{v}_{k,j} = \frac{\left(\mathbf{I} - \mathbf{H}_{k,\bar{j}}\left(\mathbf{H}_{k,\bar{j}}^H \mathbf{H}_{k,\bar{j}}\right)^{-1} \mathbf{H}_{k,\bar{j}}^H\right) \mathbf{h}_{k,j}}{\left\|\left(\mathbf{I} - \mathbf{H}_{k,\bar{j}}\left(\mathbf{H}_{k,\bar{j}}^H \mathbf{H}_{k,\bar{j}}\right)^{-1} \mathbf{H}_{k,\bar{j}}^H\right) \mathbf{h}_{k,j}\right\|_2^2}. \quad (15)$$

(15) is also employed to compute the precoder component for the target. Moreover, the suboptimal backhaul component of the precoder for each UAV $k$, $\mathbf{w}_{k,b}$ is computed as

$$\mathbf{v}_{k,b} = \frac{\left(I - \mathbf{H}_{b,\bar{k}}\left(\mathbf{H}_{b,\bar{k}}^H \mathbf{H}_{b,\bar{k}}\right)^{-1} \mathbf{H}_{b,\bar{k}}^H\right) \mathbf{h}_{k,b}}{\left\|\left(I - \mathbf{H}_{b,\bar{k}}\left(\mathbf{H}_{b,\bar{k}}^H \mathbf{H}_{b,\bar{k}}\right)^{-1} \mathbf{H}_{b,\bar{k}}^H\right) \mathbf{h}_{k,b}\right\|_2^2}, \quad (16)$$

where $\mathbf{H}_{b,\bar{k}} = [\mathbf{h}_{1,b}, ..., \mathbf{h}_{k-1,b}, \mathbf{h}_{k+1,b}, ..., \mathbf{h}_{N_{\text{Tx}},b}]$.

*2) Precoder Design:* To obtain the optimal $\mathbf{W}$ and $\mathbf{u}$ for a given position of the UAVs, similar to [15] and [16], the following procedure is adopted: first, $\mathbf{u}$ is randomly chosen and fixed. Next, since $\mathcal{P}_m$ remains non-convex due to the non-concave objective function and non-convex constraints (14b) and (14c), we begin by rewritten $\mathcal{P}_m$ as

$$\mathcal{P}'_m : \max_{\mathbf{W}} \gamma_t \quad (17a)$$

$$\text{s.t.} \left\|\begin{matrix}\tilde{\mathbf{W}}^H \mathbf{h}_j^H \\ \sigma_j\end{matrix}\right\|_2 \leq \sqrt{1+\frac{1}{\Gamma}} \mathbf{h}_j \mathbf{w}_j, \forall j \quad (17b)$$

$$\left\|\begin{matrix}\mathbf{W}_b^H \mathbf{u}\mathbf{H}_{b,k}^H \\ \sigma_k\end{matrix}\right\|_2 \leq \sqrt{1+\frac{1}{\Gamma_b}} \mathbf{u}\mathbf{H}_{b,k}\mathbf{w}_{b,k}, \forall k \quad (17c)$$

$$(14d),$$

[1] For details on the algorithm implementation, please refer to [14].

where $\mathbf{W}_b = [\mathbf{w}_{b,1}, ..., \mathbf{w}_{b,N_{\text{Tx}}}]$ and $\Gamma_b = 2^{N_{\text{ue}} \log_2(\Gamma+1)} - 1$. Next, the non-concave objective function in (17a) linearized with the first-order Taylor approximation, and the approximated optimization problem is iteratively solved via the constrained concave-convex procedure (CCCP) until a convergence criterion is attained. So, at the $i$th iteration, the following problem is solved

$$\mathcal{P}''_m : \max_{\substack{\mathbf{W}, \tau_{1,j}, \tau_{2,j}, \\ \rho_k, \hat{\rho}_{1,k}, \hat{\rho}_{2,k}}} \zeta \left(\sum_{n=1}^{N-1} \mathbf{s}^H[n]\left(\sum_{k=1}^{N_{\text{Tx}}} \frac{1}{d_{k,t}^2}\left(2\left[\mathbf{W}_k^{(i-1)}\right]^H \mathbf{A}_{k,k}\right.\right.\right.$$
$$\left.\left.\left.\times \left(\mathbf{W}_k - \mathbf{W}_k^{(i-1)}\right) + \left[\mathbf{W}_k^{(i-1)}\right]^H \mathbf{A}_{k,k} \mathbf{W}_k^{(i-1)}\right)\right)\mathbf{s}[n]\right) \quad (18a)$$

s. t. (17b), (17c), (14d),

with $\mathbf{A}_{k,k} = \mathbf{a}^*(\phi_{k,t}, \theta_{k,t})\mathbf{a}^T(\phi_{k,t}, \theta_{k,t})$. Accordingly, $\mathcal{P}''_m$ is a concave problem that can be efficiently solved by convex programming toolboxes as CVX. Finally, the optimal $i$th $\mathbf{W}$ is fixed and, $\mathbf{u}_k$ is updated via the MMSE receiver as

$$\mathbf{u}_k = \left(\mathbf{H}_{b,k}\left(\sum_{\substack{l=1 \\ l \neq i}}^{N_{\text{Tx}}} \mathbf{w}_{b,l}\mathbf{w}_{b,l}\right)\mathbf{H}_{b,k}^H + \sigma_k^2 \mathbf{I}\right)^{-1} \mathbf{H}_{b,k}\mathbf{w}_{b,k}. \quad (19)$$

### B. Tethered UAVs

For the scenario with tethered UAVs, (12) is maximized by optimizing the transmit precoder matrix $\tilde{\mathbf{W}}$ constrained to the transmit power and communications QoS requirements, which can be written as

$$\mathcal{P}_t : \max_{\tilde{\mathbf{W}}} \gamma_t \quad (20a)$$

$$\text{s. t.} \sum_{j=1}^{N_{\text{ue}}} \|\mathbf{w}_j\|^2 + \|\mathbf{w}_t\|^2 \leq P, \quad (20b)$$

$$(14b).$$

where $P$ is the total power of the system, defined as $P = P_b + N_{\text{Tx}} P_k$. Similar to the scenario with mobile UAVs, the CCCP is used to solve $\mathcal{P}_t$.

## IV. NUMERICAL RESULTS

In this section, the performance of the proposed sensing SINR maximization scheme is evaluated in terms of the sensing SINR $\gamma_t$. The simulations are performed for a set configuration of UEs spread over a square area of side $\Delta x = \Delta y$. The target is positioned at $[\frac{1}{2}\Delta x, \frac{3}{4}\Delta y]$, and the tethered UAVs at $\mathbf{r}_1 = [\frac{1}{2}\Delta x, \frac{1}{2}\Delta y, 125]$, $\mathbf{r}_2 = [\frac{1}{2}\Delta x, \frac{1}{4}\Delta y, 125]$ and $\mathbf{r}_3 = [\frac{3}{4}\Delta x, \frac{3}{4}\Delta y, 125]$. Beyond the tethered and mobile scenarios, we also consider a fixed case, where the UAVs are connected wirelessly to the BS, located at the same coordinates as the tethered case without optimizing their positions. Unless specified otherwise, the considered parameter values are depicted in Tab. I. Assumed values for frequency, bandwidths, and noise powers are extracted from [4]. Also, the fronthaul propagation and mmWave backhaul parameters are based on [13] and [11], respectively.

TABLE I
SIMULATION PARAMETERS

| Parameter | Value | Parameter | Value |
|---|---|---|---|
| $M_U$ | $16 \times 16$ | $\Gamma$ | 0 dB |
| $M_{UB}$ | 256 | $\sigma_{RCS}$ | 1 $m^2$ |
| $M_{BS}$ | 256 | $d_{min}$ | 5 m |
| $P_k$ | 30 dBm | $x_{min}$ | 0 m |
| $P_b$ | 30 dBm | $x_{max}$ | 500 m |
| $N_{Tx}$ | 3 | $y_{min}$ | 0 m |
| $N_{ue}$ | 20 | $y_{max}$ | 500 m |
| $N$ | 64 | $z_{min}$ | 20 m |
| $N_0/BW$ | -174 dBm/Hz | $z_{max}$ | 200 m |

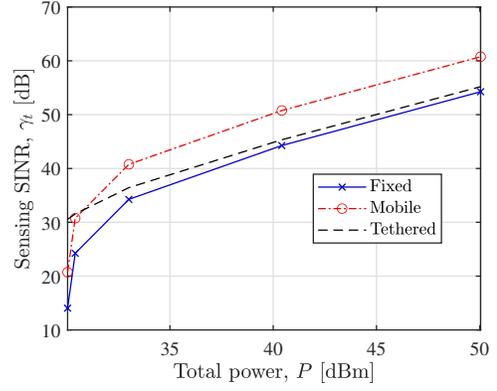

Fig. 4. Sensing SINR $\gamma_t$ vs. the total power $P$, for the three considered cases.

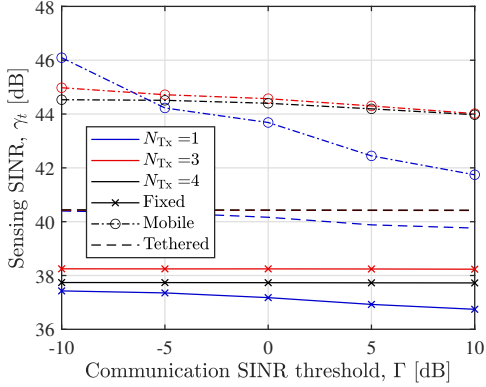

Fig. 3. Sensing SINR $\gamma_t$ vs. the communication SINR threshold, $\Gamma$, for $N_{Tx}=1, 3$ and 4 with fixed total system power, $P = 40$ dBm.

Fig. 3 shows the sensing SINR $\gamma_t$ vs. the communication SINR threshold, $\Gamma$, for a different number of transmit UAVs $N_{Tx}$. For all scenarios, the total power is assumed $P=40$ dBm, split between the BS and UAVs as $P_b=P_k=P/(N_{Tx}+1)$, $\forall k$ for the fixed and mobile cases. The mobile case outperforms the fixed and tethered case, indicating the advantage of employing the proposed position control algorithm. When only one transmit UAV is employed, the performance degrades for all cases. This is expected, since given the size of the area of the system, it is difficult for only one UAV to attend the strict QoS requirements of the UEs. For the fixed case, increasing the number of UAVs to three improves the sensing SINR. However, with four transmit UAVs, a degradation of performance is observed. This occurs because, for the fixed case, the UAVs use all the available power, which is further fractioned as more UAVs are present. For the scenario with tethered UAVs, the performance is similar for a higher number of transmitting UAVs. This happens because, given the centralized power allocation, a specific transmit UAV may have more power allocated to it if the communication link between the UAV and a UE is the strongest. Finally, the same tendency can be seen for the mobile case. However, as $\Gamma$ increases, the case with three UAVs has a steeper slope than the case with four UAVs. Specifically, this is more pronounced for the mobile case since the positioning of the transmit UAVs can be leveraged to reduce the effect on the sensing SINR, differently from the other cases.

Fig. 4 illustrates the sensing SINR $\gamma_t$ vs. the total power of the system, $P$, with $P_b$ fixed. Observe that the mobile case always outperforms the fixed one, and it surpasses the performance of the tethered case for $P$ higher than 30 dBm. This indicates the advantage of employing the position protocol algorithm. Note that, for the tethered case, the sensing SINR grows linearly with the total power of the system, and presents better performance than the other cases for a low value of $P$. This is expected, since, at low $P$ the total power of the system mostly comprises the available power at the BS, $P_b$. For the tethered case, this power can be employed by transmit UAVs, whereas, for the mobile and fixed scenarios, $P_b$ does not present any effect on the sensing SINR. However, as the total power increases, the influence of $P_b$ in the tethered case is reduced, with its performance approaching that of the fixed one.

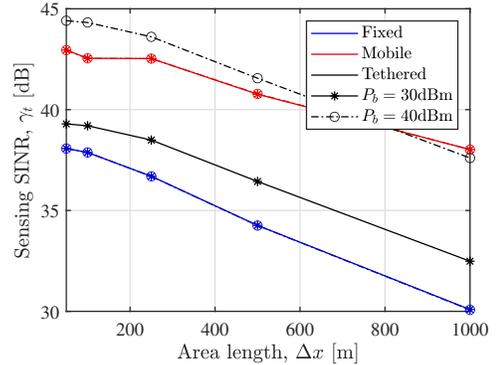

Fig. 5. Sensing SINR $\gamma_t$ vs. the area length, $\Delta x$ for $P_b = 30$ and 40 dBm.

Fig. 5 shows the sensing SINR $\gamma_t$ vs. the area length $\Delta x$ for $P_b = 30$ dBm and $P_b = 40$ dBm, with $P_k$ fixed. Note that, validating the results obtained in Fig. 4, by increasing $P_b$, no variation in performance is observed for the fixed or mobile cases, since $P_b$ is just employed for the backhaul link, and does not affect the attained sensing SINR. On the other hand, since the total power available for transmission at the UAVs increases with the total power of the BS for the tethered case, it can outperform the mobile case for higher values of $P_b$. Moreover, note that the performance of the tethered case decreases faster than the mobile case as the area length increases. This implies that, in larger areas of service the mobile case outperforms the tethered case, even in high system power conditions. This occurs because, in larger areas, the position control algorithm can ensure that the UAVs take more beneficial positions toward the target while maintaining the QoS for the UEs.

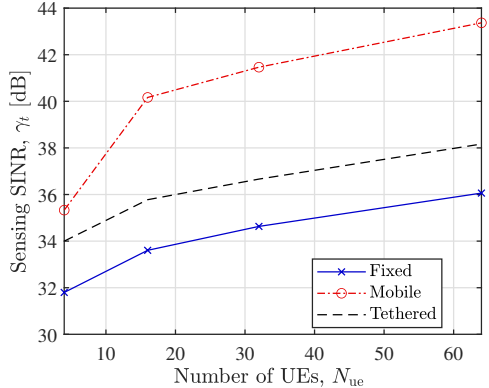

Fig. 6. Sensing SINR $\gamma_t$ vs. number of UEs $N_{\text{ue}}$, for the fixed, mobile, tethered cases.

Finally, Fig. 6 shows the sensing SINR $\gamma_t$ vs. the number of UEs $N_{\text{ue}}$. Note that, increasing the number of UEs improves the sensing performance of the system. To explain this, note that in (13), the precoding matrix $\mathbf{W}_k$ has a rank of $\min\{N_{\text{ue}} + 1, M_U\}$. Thus, increasing $N_{\text{ue}}$ provides a gain in $\gamma_t$. Also, note that this gain is more pronounced when $N_{\text{ue}}$ goes from $N_{\text{ue}}=4$ to $N_{\text{ue}}=16$. This occurs because, even though there are more degrees of freedom available to aid in the sensing of the target as the number of UEs increases, these are first utilized to meet the communications constraints. Also, as the number of UEs increases, more power has to be allocated for communication, which also influences on the less noticeable sensing performance improvement.

## V. CONCLUSIONS

We investigated the use of multiple UAVs in a cell-free mMIMO architecture for ISAC with a dedicated signal for sensing. Specifically, three different deployments were considered for the UAVs, namely, mobile UAVs, tethered UAVs, and the fixed case, where the position of the UAVs was not optimized. From the results, we showed that when the power available for the UAVs is low, the tethered scenario provides a better performance in terms of sensing SINR. However, as the power available for the UAVs increases, the proposed positioning control algorithm is capable of achieving better sensing SINR performance. Moreover, increasing the number of UEs in the system can improve the sensing SINR, since the communication signals intended for the UEs can be leveraged for sensing. Finally, although the optimal number of UAVs deployed in the network needs to be carefully assessed, employing a position control algorithm for multiple mobile UAVs is capable of overcoming limitations in terms of power and surpasses the performance of tethered UAVs in most cases.

### A. Future Works

Potential future works includes the sequential UAV movement problem for UAV position control, the evaluation of the energy consumption due to the movement of the UAVs, as well as the comparison to cellular configurations.

ACKNOWLEDGEMENT

This research was supported by the Research Council of Finland (former Academy of Finland) 6G Flagship Programme (Grant Number: 346208).